\documentclass[conference]{IEEEtran}

\usepackage{graphicx}
\usepackage{amsmath,amssymb,amsfonts}
\usepackage{cite}
\usepackage{booktabs}
\usepackage{array}
\usepackage{algorithmic}
\usepackage{hyperref}
\usepackage{tabularx}
\usepackage{caption}
\usepackage{tikz}
\usetikzlibrary{calc}
\usepackage{longtable}
\usepackage{tabularray}
\usepackage{microtype}
\usepackage[all]{nowidow}
\begin{document}

\title{Acoustic Analysis of Uneven Blade Spacing and Toroidal Geometry for Reducing Propeller Annoyance\\\vspace{1ex}\normalsize{Manuscript completed: October 6, 2023}\\\normalsize{Revised: \today}}

\author{%
	 \IEEEauthorblockN{%
		Nikhil Vijay\IEEEauthorrefmark{1},
		Will C. Forte\IEEEauthorrefmark{1},
		Ishan Gajjar\IEEEauthorrefmark{1},
		Sarvesh Patham\IEEEauthorrefmark{1}%
	}%
	\IEEEauthorblockN{%
		Syon Gupta\IEEEauthorrefmark{1},
		Sahil Shah\IEEEauthorrefmark{1},
		Prathamesh Trivedi\IEEEauthorrefmark{1},
		Rishit Arora\IEEEauthorrefmark{1}%
	}%
    \IEEEauthorblockA{\IEEEauthorrefmark{1}The Academy for Mathematics, Science, and Engineering, Rockaway, NJ, USA\\
    Email: \{nikhilvijayengineering, willcforte, ishangajjar10, sarveshpatham, sahil.sa.shah, thunderpra, rishitarora2005\}@gmail.com}
}

\maketitle

\begin{abstract}
Unmanned aerial vehicles (UAVs) are becoming more commonly used in populated areas, raising concerns about noise pollution generated from their propellers. This study investigates the acoustic performance of unconventional propeller designs, specifically toroidal and uneven-blade spaced propellers, for their potential in reducing psychoacoustic annoyance. Our experimental results show that these designs noticeably reduced acoustic characteristics associated with noise annoyance.
\end{abstract}

\begin{IEEEkeywords}
Broadband noise, tonal noise, vortices, noise pollution, airfoil, CFD simulation, power spectral density (PSD), uneven blade spacing, toroidal propeller.
\end{IEEEkeywords}

\section{Introduction}

The recent boom in drone usage allows such technology to be used in a
variety of consumer and industrial applications, including surveillance,
delivery, agriculture, filmmaking, rescue, and further applications for
land-based and aquatic drones {[}1{]}-{[}2{]}. As multirotor drones and
unmanned aerial vehicles (UAVs) become more common within populous
areas, the environmental problems resulting from the use of such
vehicles become apparent, especially noise pollution. Research by
Schäffer et al. found the psychoacoustic annoyance of drones to even
exceed that of airplanes or automotives at similar sound pressure
levels. This, combined with the fact that the public is still forming
opinions on the acceptability of larger-scale UAV usage in their
communities, makes the development of high efficiency, low-annoyance
propellers necessary for the future of civil UAV applications {[}4{]}.

Propeller noise can be separated into two main categories: tonal and
broadband noise. Tonal noise is created by the harmonics of the blade
passage frequency (BPF; rotational frequency multiplied by the number of
blades), while broadband noise is created by the formation of vortices
at the tips and trailing edge of the blade, resulting in excessive
turbulence and, consequently, heightened levels of acoustic output
{[}4{]}. Uneven blade spacing has shown promise in reducing perceived
tonal noise by spreading out the frequencies over a wider range,
especially in the lower ranges {[}5{]}.

Such irritation can be attributed to the special acoustic
characteristics of propellers, particularly pure tones (sinusoidal
waveforms) and high-frequency broadband noise {[}4{]}. Pure-tone effects
can be more irritating to humans than random noise distributed more
equally along different noise frequencies. Kryter and Pearsons
illustrate that sound consisting of a pure tone superimposed on
background noise can be made to sound less noisy by dispersing the
energy of the tone over several discrete frequencies {[}5{]}.
Low-annoyance propeller design should aim to minimize pure-tone effects
as well as reduce the overall broadband noise level in order to be
effective.

\section{Background}

\textbf{Conventional Propellers: }Traditionally, propellers employ
linear blades that are spaced at even intervals around the propeller hub
to evenly distribute mass. The blades are designed to create an area of
high pressure under each blade when spinning, thus generating lift.
Varying rotational speeds allows for different amounts of vertical lift
to be generated by the propeller, with faster rotation leading to
greater lift {[}4{]}.

Traditional propellers, however, emit large amounts of broadband noise,
as the vortices formed at the tips of each blade create high turbulence
{[}3{]}.

\textbf{Toroidal Propellers: }The toroidal propeller minimizes this
effect with blades whose tips ``fold over'' into the adjacent blade,
thus reducing the impact of vortices {[}5{]}. With the effect of
vortices minimized, the noise emitted from the propeller at the
frequency range to which humans are most perceptive is also minimized,
thus decreasing the perceived loudness.

\textbf{Uneven Blade Spacing: }The application of uneven blade spacing
in drone design, as used by companies like Zipline, offers a novel
solution to mitigate tonal noise. Uneven blade spacing disrupts the
alignment of pressure pulses generated by the rotor blades, diffusing
the concentrated tonal noise into a broader frequency range, thus also
reducing the perceived loudness {[}1{]}. While the integration of uneven
blade spacing does necessitate careful navigation of associated
engineering challenges such as the maintenance of aerodynamic efficiency
and flight stability, it is an avenue that holds promising potential for
decreasing sound annoyance.

\section{Motivations \& Proposal}

To collect rigorous, useful experimental data for the different
propeller designs, they must adhere to similar constraints involving
properties including size, design methodology, and material. They must
also be evaluated for both noise and thrust performance.

\textbf{Materials \& Fabrication: }Additive manufacturing, commonly
implemented through 3D printing, allows for the rapid and effective
development of drone propellers {[}7{]}. Due to the relatively low cost
and fast production speed of 3D printing compared to traditional
propeller manufacturing methods such as injection molding and metal
machining, prototypes can be produced efficiently. While 3D printed
propellers generally tend to have less desirable performance than
commercially manufactured parts {[}8{]}, such limitations are negligible
for the purposes of this paper as all propellers are fabricated using
the same method.

\begin{table}[h]
    \caption{Tested Propellers}
        \label{tab:tested_propellers}
    \centering
    \begin{tblr}{colspec={X[3,l] X[5,l] X[2,c]}, hlines}
        \textbf{Propeller Type} & \textbf{Role in Study} & \textbf{Mass (g)} \\
        Type A - 3-blade conventional propeller & Control for 3-blade propellers & 6 \\
        Type B - 3-loop toroidal propeller & Standard toroidal propeller & 11 \\
        Type C - 2-loop toroidal propeller with uneven blade spacing and counterweight & Used to see the effect of uneven blade spacing on toroidal propeller & 13 \\
        Type D - 6-blade conventional propeller & Control for 6-blade propellers & 10 \\
        Type E - 6-blade propeller with uneven blade spacing & Used to see the effect of uneven blade spacing on 6-blade propeller & 10 \\
    \end{tblr}
\end{table}

\textbf{\textnormal{Design Methodology: }}\textnormal{Autodesk Fusion 360 was
used to }\textnormal{design}\textnormal{ all propellers from the same custom
airfoil profile. Each model has a diameter of 6 inches.}\textnormal{ Five
designs were created, }\textnormal{as }\textnormal{seen in Table I. Two
conventional propellers are included, Type A, with three blades, and
Type D, with six blades.}

\textbf{\textnormal{Unconventional Propellers: }}\textnormal{Three
unconventional designs were created for testing and comparison to the
conventional propellers}\textnormal{ (types B, C, and E)}\textnormal{. }

\textnormal{Type B, the 3-loop toroidal, utilizes similar airfoil profiles
as the conventional propellers, but the tips of each blade are swept
into each other, creating three loops. }

\textnormal{Type C, the 2-loop toroidal propeller with uneven blade spacing
and counterweight,}\textnormal{ involves two adjacent toroidal
}\textnormal{``}\textnormal{loops}\textnormal{''}\textnormal{ at a
71.6}\textnormal{°}\textnormal{ angle (relevant to a line drawn from the
propeller hub through the center of each loop to its outermost edge)
with an airfoil opposite to the blades serving as a counterweight. }

\textnormal{Type E, the 6-blade with uneven blade spacing, also uses the
airfoil profiles of the conventional propellers, but their spacing is
offset. }

\textnormal{As previously established, a toroidal }\textnormal{loop}\textnormal{
design}\textnormal{ }\textnormal{reduces }\textnormal{broadband noise }\textnormal{by
eliminated}\textnormal{ tip vortices, }\textnormal{while}\textnormal{
}\textnormal{unevenly spaced blades, in conjunction with a counterweight,
can reduce tonal noise}\textnormal{. In isolation, each design is effective
at minimizing a specific type of noise }\textnormal{--}\textnormal{ broadband or
tonal }\textnormal{--}\textnormal{ }\textnormal{emitted from}\textnormal{ a propeller,
but has }\textnormal{little}\textnormal{ }\textnormal{impact on the other,
}\textnormal{severely impeding the ability of each to minimize noise
pollution. We propose}\textnormal{ that combining these two designs will
lead to a greater net decrease in noise output from drone propellers,
reducing the auditory }\textnormal{burden}\textnormal{ such propellers
}\textnormal{inflict}\textnormal{ on humans.}

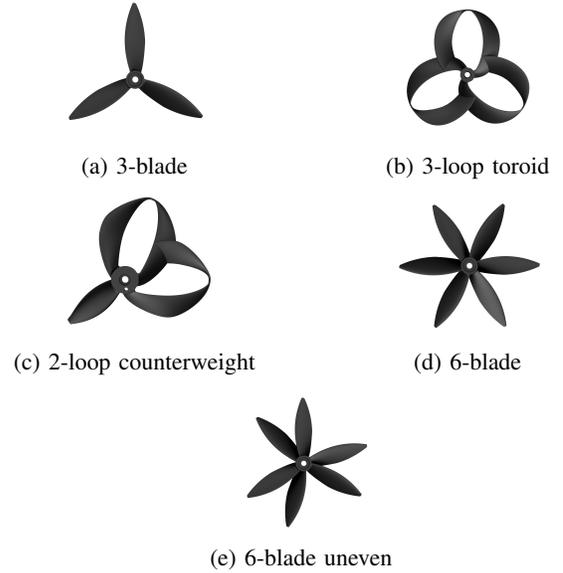
\begin{figure}[ht]
    \centering
    \begin{tikzpicture}[every node/.style={inner sep=0}]
    	\def\imgwidth{0.48\linewidth}
    	\def\imgheight{2.2cm}
    	\def\xoffset{0.25\linewidth}
    	\def\ysep{2.6cm}
    	
    	% First row
    	\node[anchor=center] (img1) at (-\xoffset, 0) {\includegraphics[width=\imgwidth, height=\imgheight, keepaspectratio]{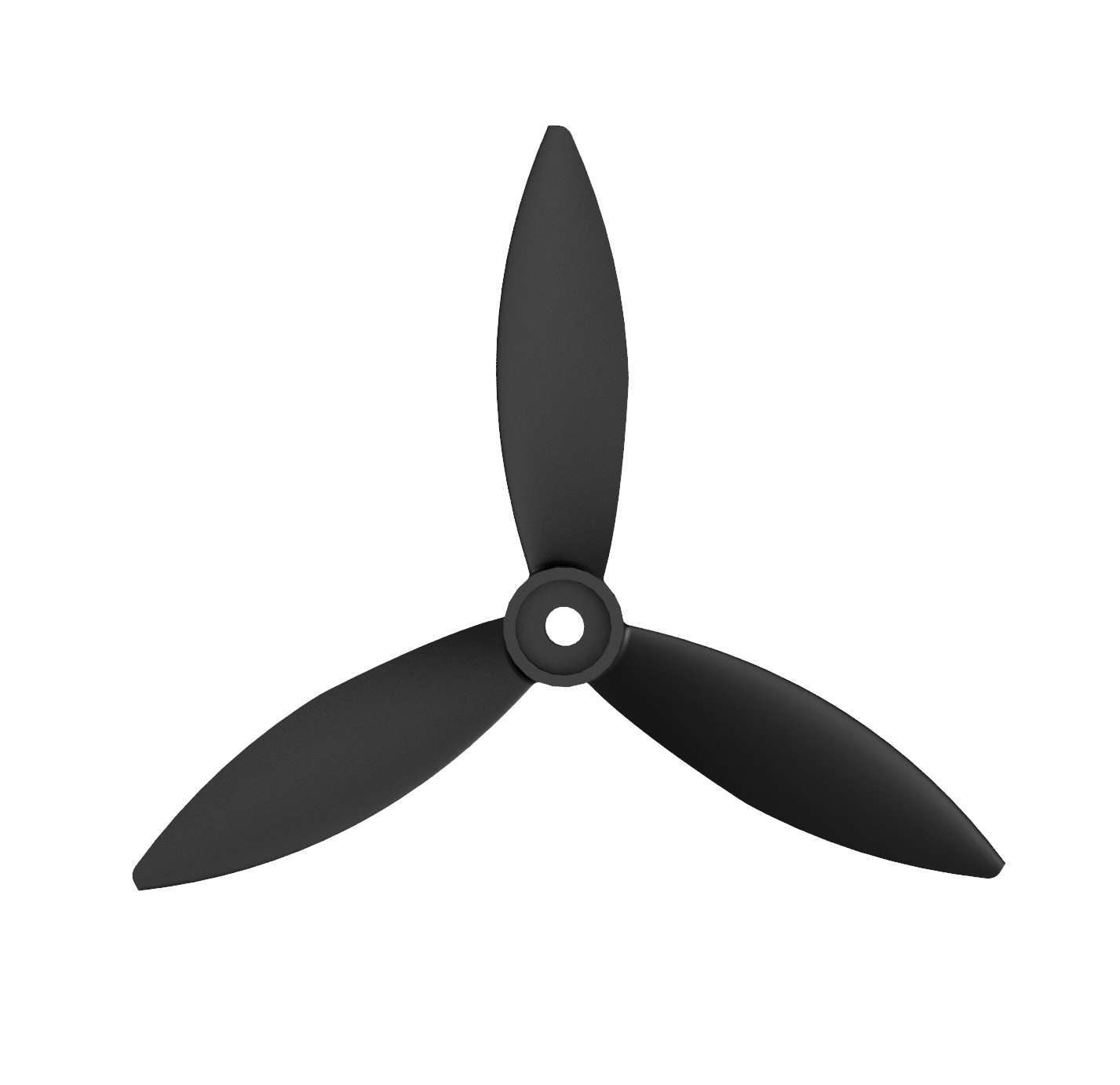}};
    	\node[anchor=center] (img2) at ( \xoffset, 0) {\includegraphics[width=\imgwidth, height=\imgheight, keepaspectratio]{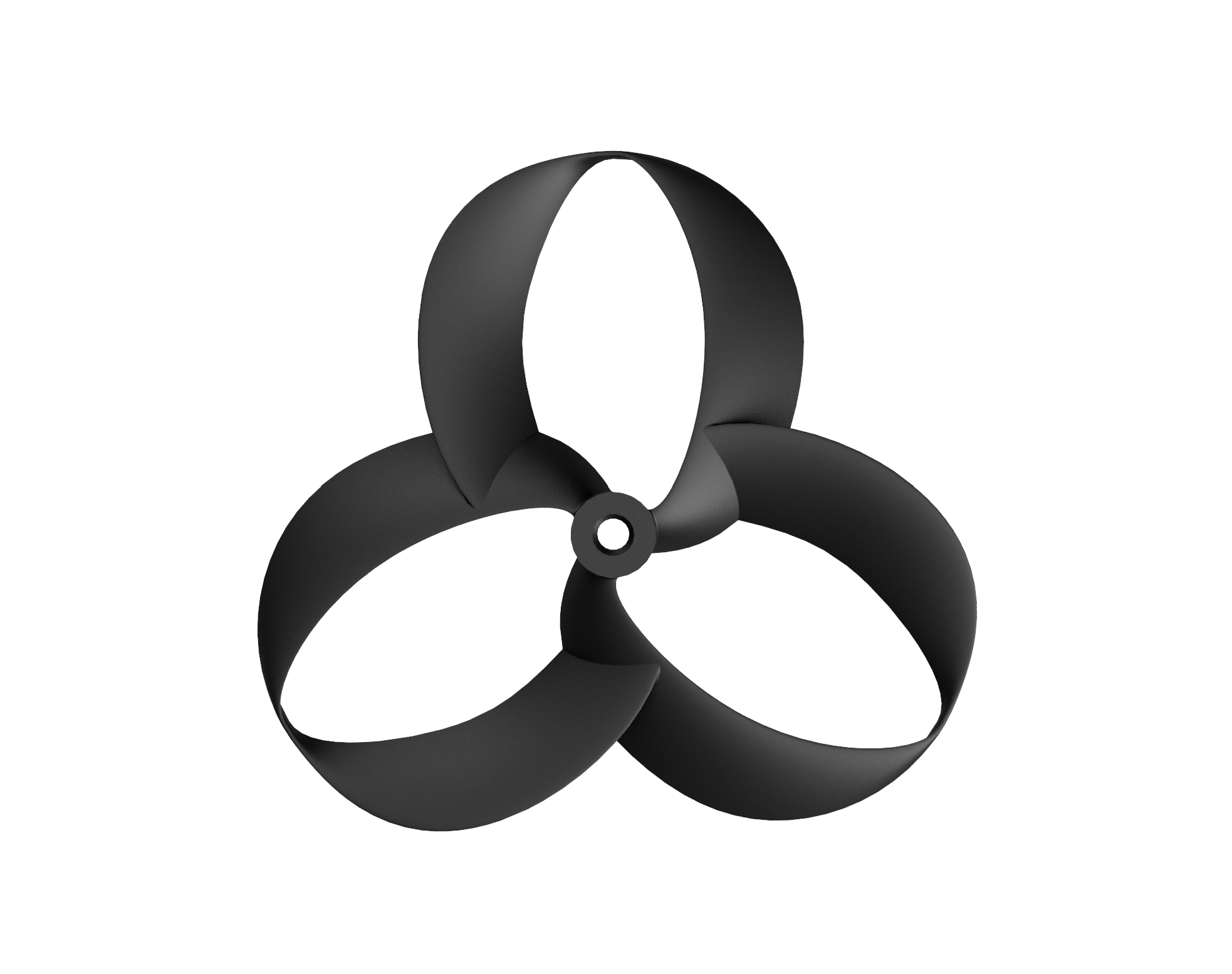}};
    	
    	% Second row
    	\node[anchor=center] (img3) at (-\xoffset, -\ysep) {\includegraphics[width=\imgwidth, height=\imgheight, keepaspectratio]{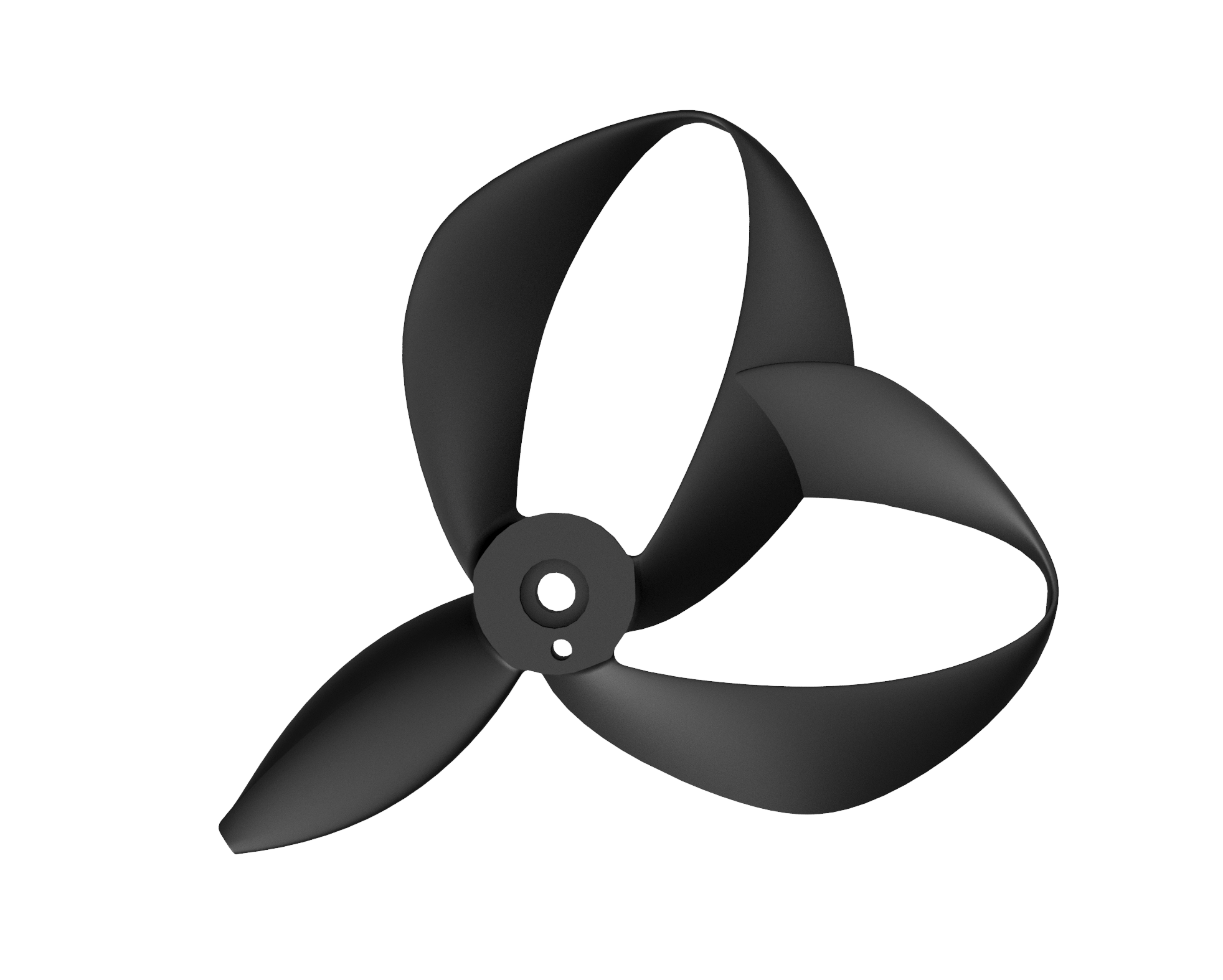}};
    	\node[anchor=center] (img4) at ( \xoffset, -\ysep) {\includegraphics[width=\imgwidth, height=\imgheight, keepaspectratio]{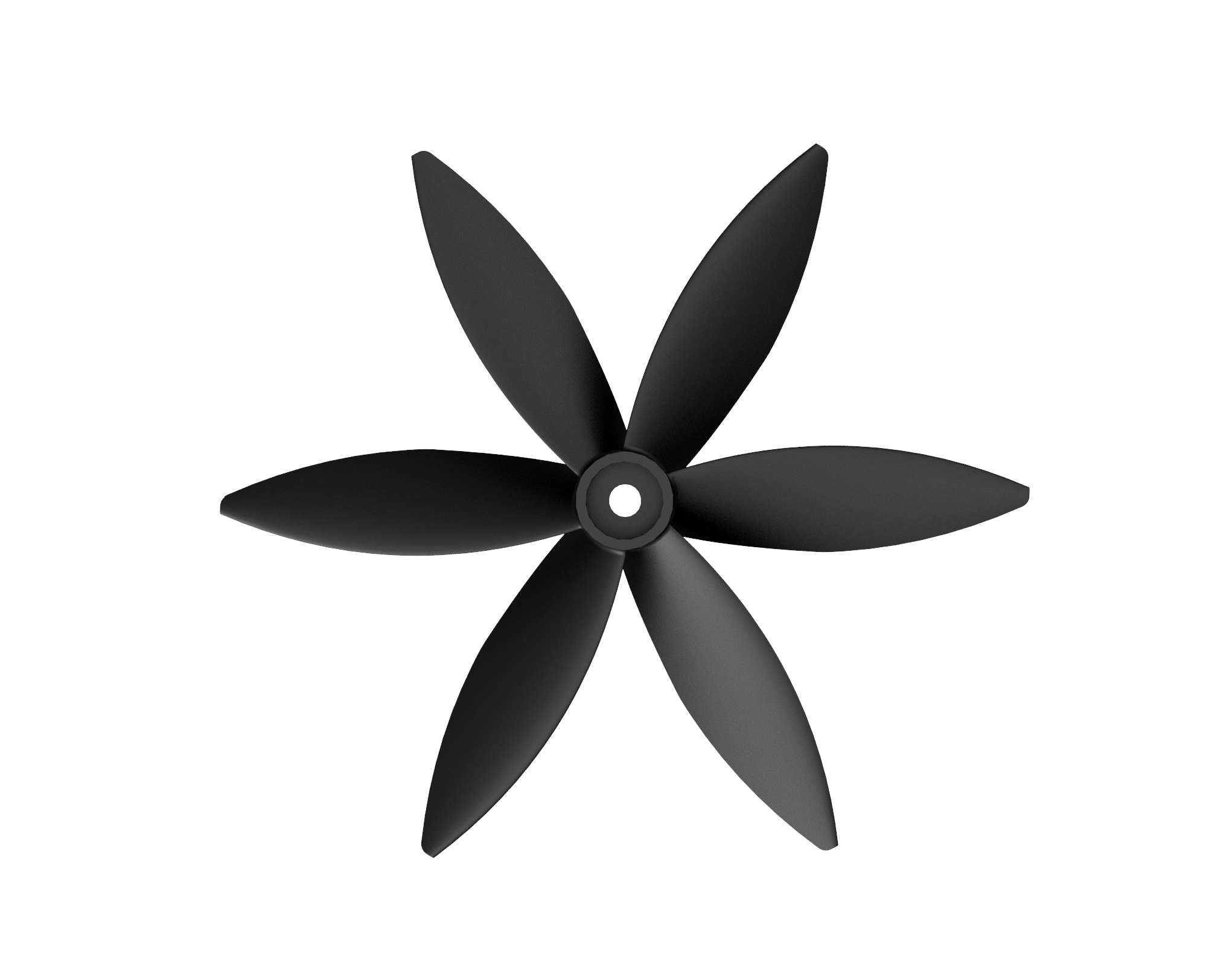}};
    	
    	% Third row
    	\node[anchor=center] (img5) at (0, {-2*\ysep}) {\includegraphics[width=\imgwidth, height=\imgheight, keepaspectratio]{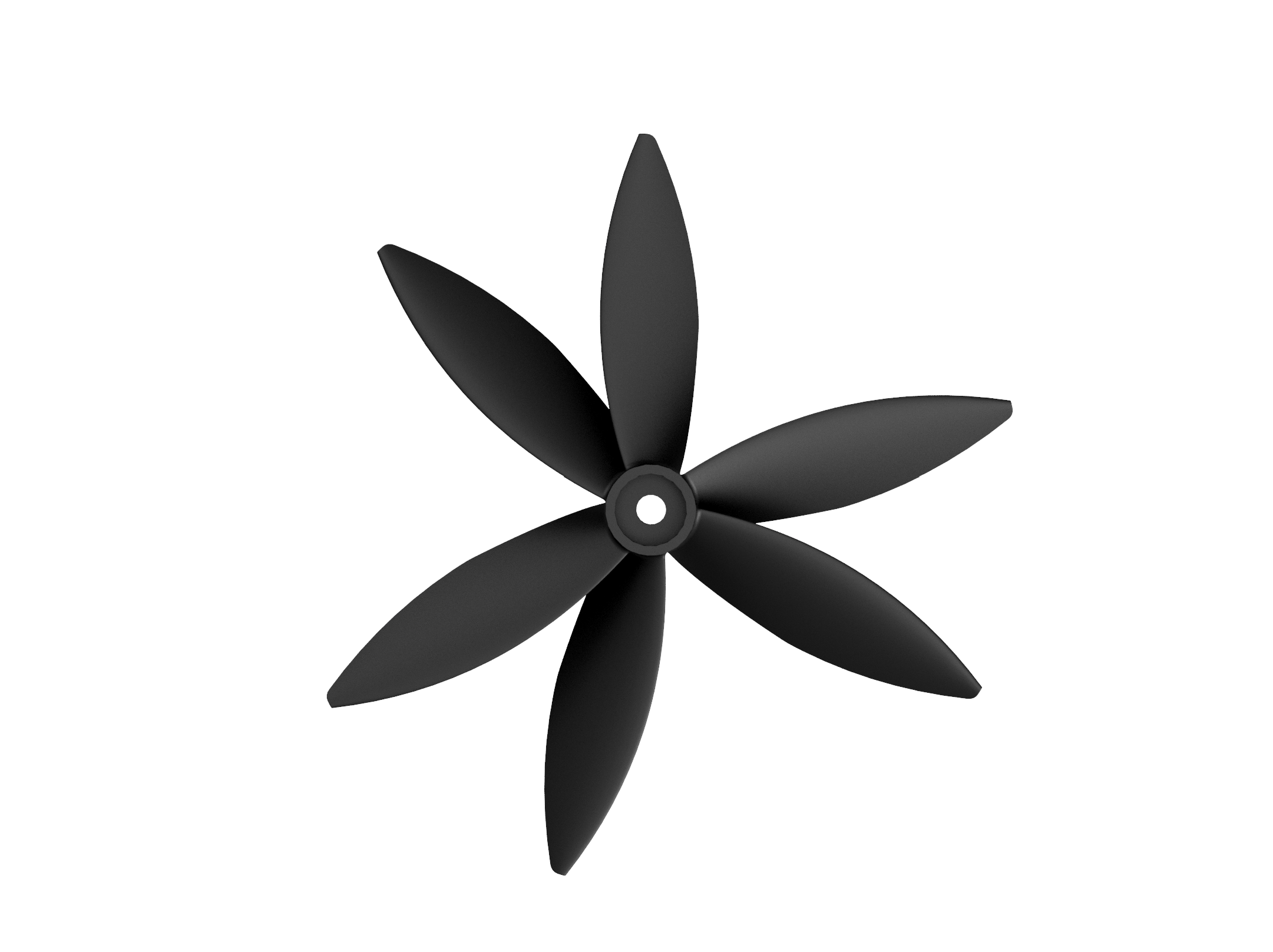}};
    	
    	% Labels
    	\node[anchor=north, font=\small] at ($(img1.south) + (0,-0.1)$) {(a) 3-blade};
    	\node[anchor=north, font=\small] at ($(img2.south) + (0,-0.1)$) {(b) 3-loop toroid};
    	\node[anchor=north, font=\small] at ($(img3.south) + (0,-0.1)$) {(c) 2-loop counterweight};
    	\node[anchor=north, font=\small] at ($(img4.south) + (0,-0.1)$) {(d) 6-blade};
    	\node[anchor=north, font=\small] at ($(img5.south) + (0,-0.1)$) {(e) 6-blade uneven};
    \end{tikzpicture}
    \caption{3D models of our propeller types, three of which (b,c,e) were tested using physical prototypes.}
    \label{fig:unconventional_propellers}
\end{figure}

\section{Experimental Procedure}

Before experimentally testing the designs, basic efficiency tests were
simulated in order to determine expected performances for each
propeller. Using Ansys Fluent, a computational fluid dynamics (CFD)
simulation software, thrust, force, and velocity analyses were run on
all propellers to compare their aerodynamic performance. Part of the
simulation parameters from {[}7{]} were used. Although thrust was tested
over a range of angular velocities, a value of 6,000 rotations per
minute (RPM) was used for visual CFD analysis (Fig. 2-4).

\begin{table}[h]
    \caption{CFD Parameters}
    \label{tab:cfd_parameters}
    \centering
    \begin{tblr}{colspec={X[3,l] X[3,l]}, hlines}
        \textbf{Parameter} & \textbf{Value} \\
        Angular velocity range & [1, 21] kRPM \\
        Gravity & 9.81 m/s\textsuperscript{2} \\
        Time & Transient \\
        Viscous model & K-epsilon (realizable) \\
        Near wall treatment & Scalable wall function \\
        Inlet air velocity & 0 m/s \\
        Iterations per time step & 80 \\
        Time step size & 0.125 \\
        Number of time steps & 30 \\
    \end{tblr}
\end{table}

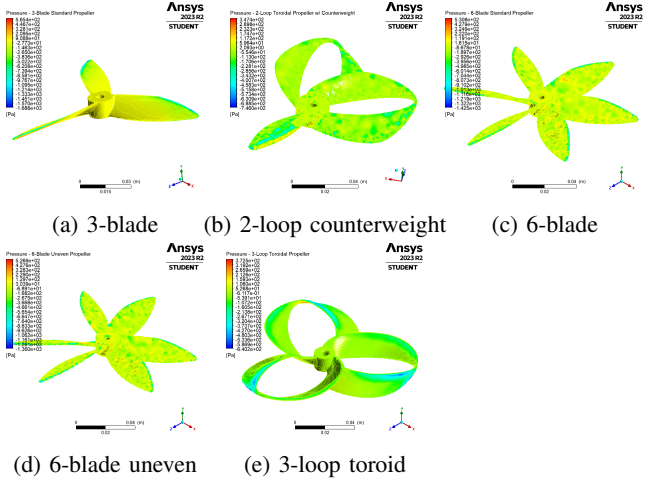
\begin{figure}[ht]
    \centering
    \begin{tikzpicture}[every node/.style={inner sep=0}]
    	\def\imgwidth{0.3\linewidth}  % Approximate 1/3 of text width
    	\def\xsep{0.33\linewidth}     % Horizontal spacing
    	\def\ysep{3.2cm}              % Vertical spacing between rows
    	
    	% First row
    	\node[anchor=north west] (img1) at (0, 0) {\includegraphics[width=\imgwidth]{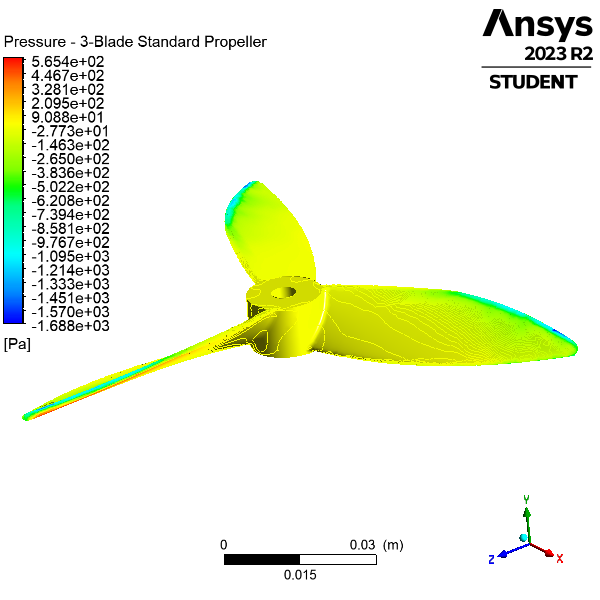}};
    	\node[anchor=north west] (img2) at (\xsep, 0) {\includegraphics[width=\imgwidth]{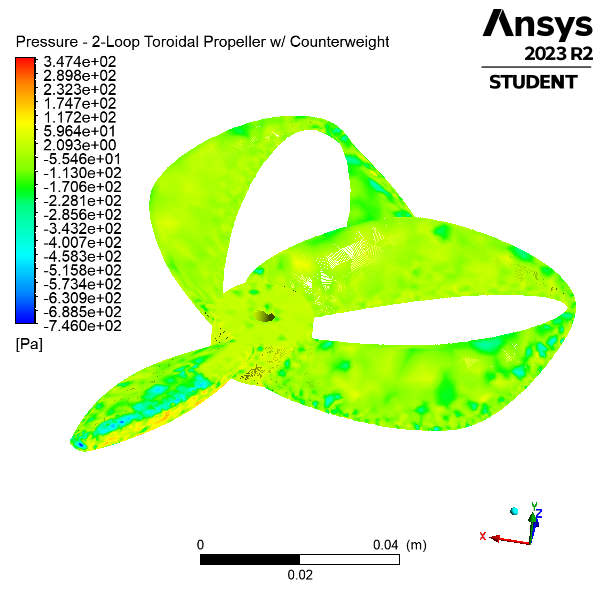}};
    	\node[anchor=north west] (img3) at ({2*\xsep}, 0) {\includegraphics[width=\imgwidth]{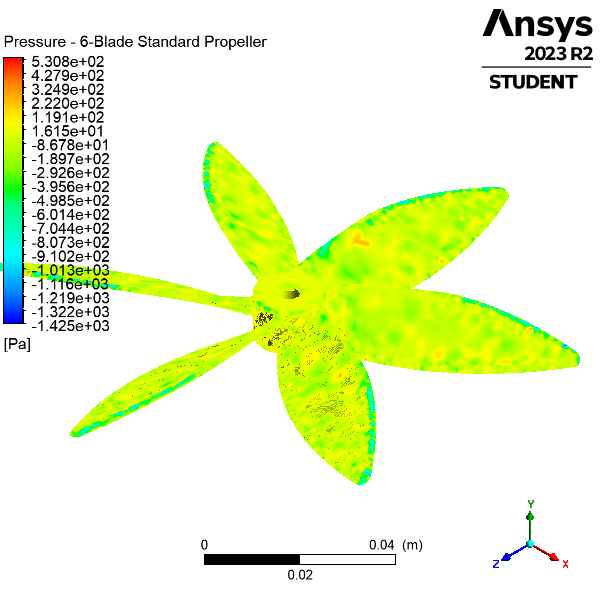}};
    	
    	% Second row
    	\node[anchor=north west] (img4) at (0, -\ysep) {\includegraphics[width=\imgwidth]{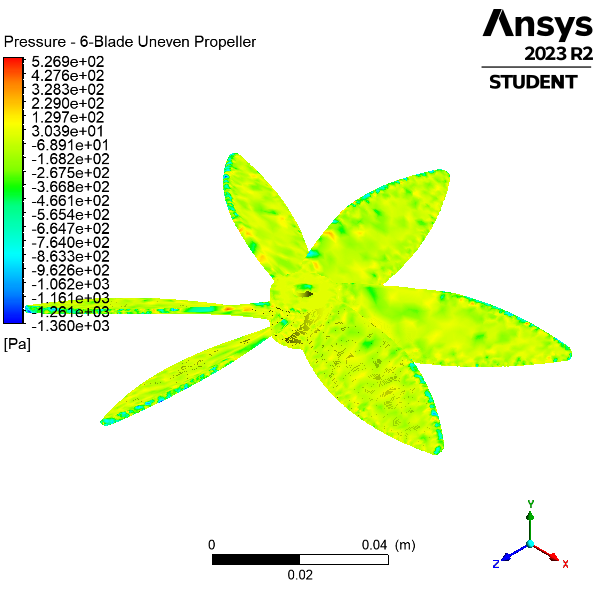}};
    	\node[anchor=north west] (img5) at (\xsep, -\ysep) {\includegraphics[width=\imgwidth]{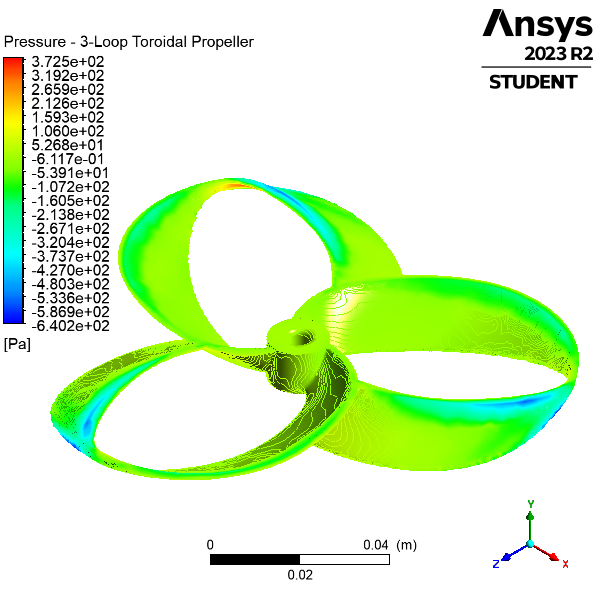}};
    	%\node[anchor=north west] (img6) at ({2*\xsep}, -\ysep) {\includegraphics[width=\imgwidth]{figures/docs_export/image6.png}};
    	
    	% Labels below images
    	\node[anchor=north, font=\small] at ($(img1.south) + (0,-0.2)$) {(a) 3-blade};
    	\node[anchor=north, font=\small] at ($(img2.south) + (0,-0.2)$) {(b) 2-loop counterweight};
    	\node[anchor=north, font=\small] at ($(img3.south) + (0,-0.2)$) {(c) 6-blade};
    	\node[anchor=north, font=\small] at ($(img4.south) + (0,-0.2)$) {(d) 6-blade uneven};
    	\node[anchor=north, font=\small] at ($(img5.south) + (0,-0.2)$) {(e) 3-loop toroid};
    	%\node[anchor=north, font=\small] at ($(img6.south) + (0,-0.2)$) {(f)};
    \end{tikzpicture}
    \caption{CFD simulation of surface pressure.}
    \label{fig:surface_pressure}
\end{figure}

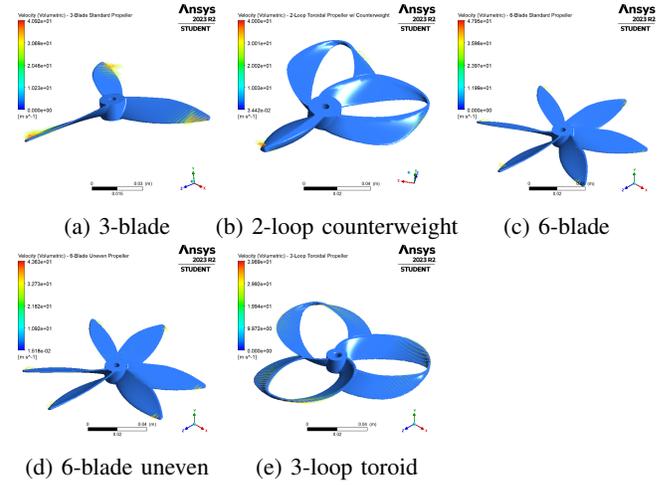
\begin{figure}[ht]
    \centering
    \begin{tikzpicture}[every node/.style={inner sep=0}]
    	\def\imgwidth{0.3\linewidth}  % Approximate 1/3 of text width
    	\def\xsep{0.33\linewidth}     % Horizontal spacing
    	\def\ysep{3.2cm}              % Vertical spacing between rows
    	
    	% First row
    	\node[anchor=north west] (img1) at (0, 0) {\includegraphics[width=\imgwidth]{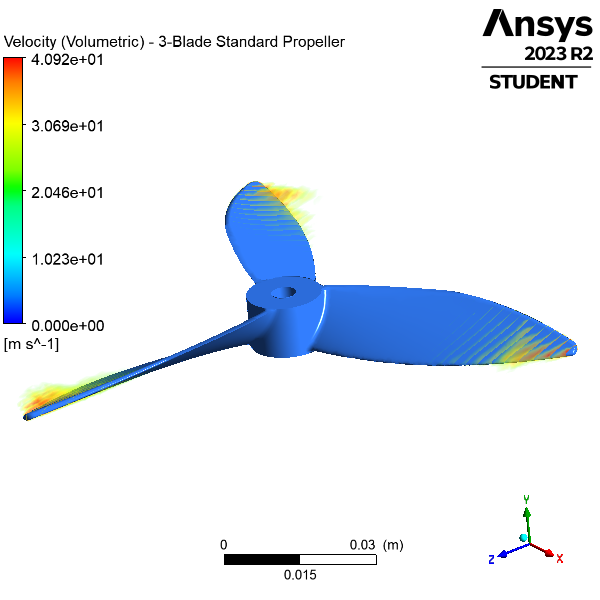}};
    	\node[anchor=north west] (img2) at (\xsep, 0) {\includegraphics[width=\imgwidth]{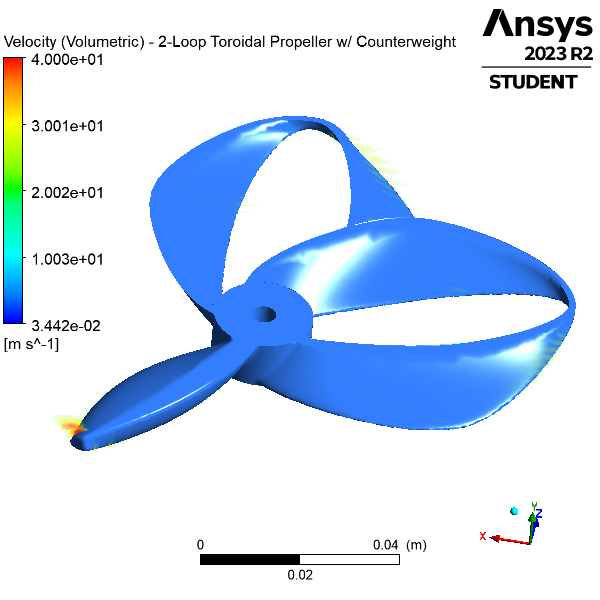}};
    	\node[anchor=north west] (img3) at ({2*\xsep}, 0) {\includegraphics[width=\imgwidth]{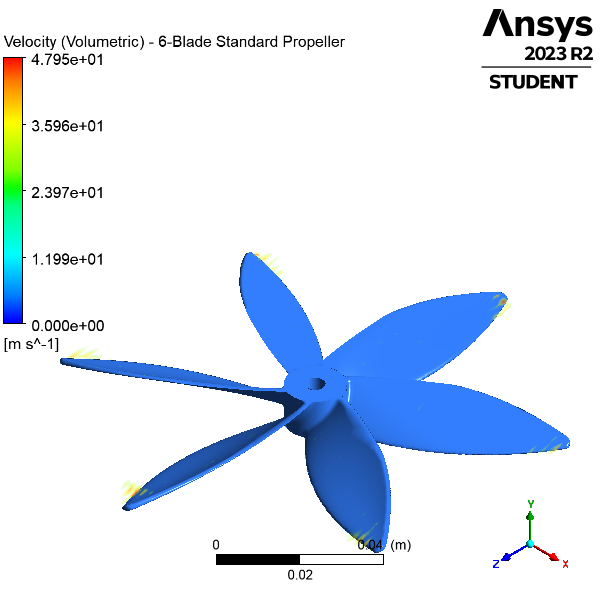}};
    	
    	% Second row
    	\node[anchor=north west] (img4) at (0, -\ysep) {\includegraphics[width=\imgwidth]{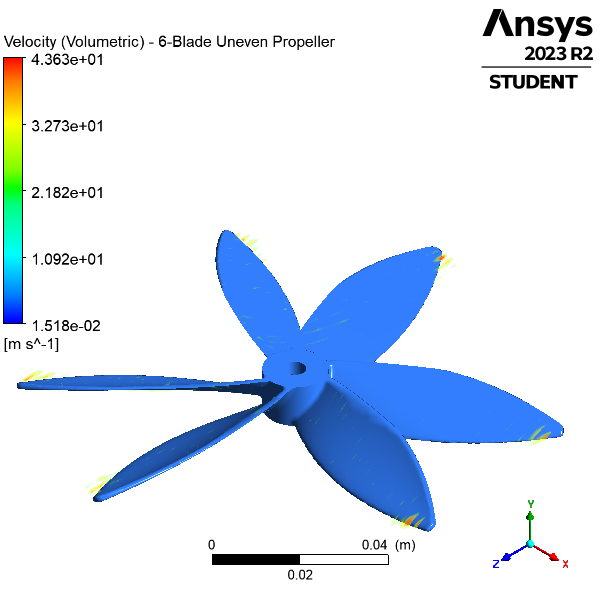}};
    	\node[anchor=north west] (img5) at (\xsep, -\ysep) {\includegraphics[width=\imgwidth]{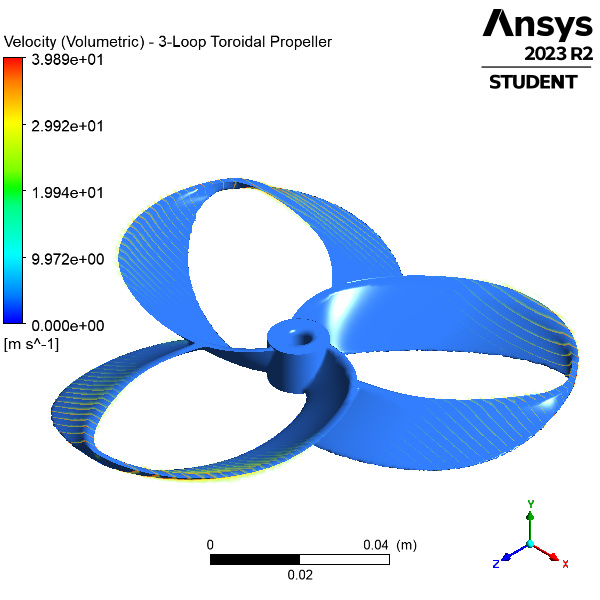}};
    	%\node[anchor=north west] (img6) at ({2*\xsep}, -\ysep) {\includegraphics[width=\imgwidth]{figures/docs_export/image6.png}};
    	
    	% Labels below images
    	\node[anchor=north, font=\small] at ($(img1.south) + (0,-0.2)$) {(a) 3-blade};
    	\node[anchor=north, font=\small] at ($(img2.south) + (0,-0.2)$) {(b) 2-loop counterweight};
    	\node[anchor=north, font=\small] at ($(img3.south) + (0,-0.2)$) {(c) 6-blade};
    	\node[anchor=north, font=\small] at ($(img4.south) + (0,-0.2)$) {(d) 6-blade uneven};
    	\node[anchor=north, font=\small] at ($(img5.south) + (0,-0.2)$) {(e) 3-loop toroid};
    	%\node[anchor=north, font=\small] at ($(img6.south) + (0,-0.2)$) {(f)};
    \end{tikzpicture}
    \caption{CFD simulation of air velocity, represented volumetrically. Areas of high velocity at the ends of blades are tip vortices.}
    \label{fig:air_velocity}
\end{figure}

\begin{figure}[ht]
    \centering
    \includegraphics[width=\linewidth, keepaspectratio]{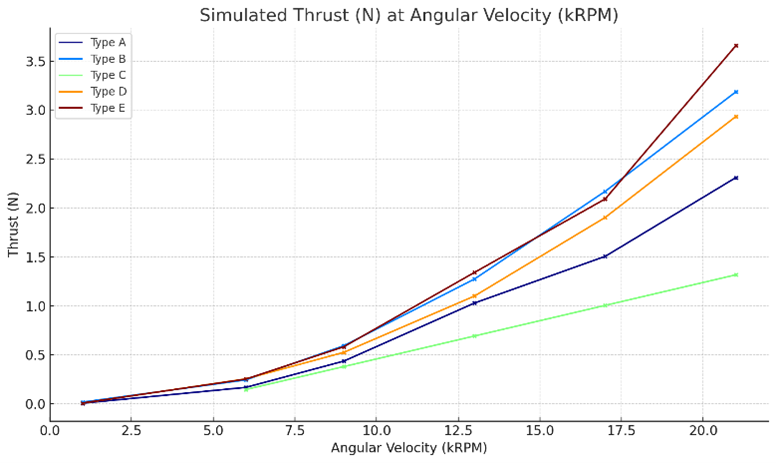}
    \caption{Simulated thrust values by angular velocity for all propellers.}
    \label{fig:thrust_values}
\end{figure}

\textnormal{With CFD analysis, }\textnormal{the aerodynamic performance}\textnormal{
of the propellers}\textnormal{ }\textnormal{can}\textnormal{ be predicted
}\textnormal{prior to physical testing}\textnormal{.
}\textnormal{P}\textnormal{ropellers of}\textnormal{ }\textnormal{types B, D, and E were
simulated to }\textnormal{generate thrust}\textnormal{ within 0.01 Newtons of
each other (Fig. 4), indicating that }\textnormal{unevenly spaced}\textnormal{
and toroidal designs do not significantly sacrifice performance with
respect to conventional designs}\textnormal{.}

\textnormal{CFD analysis can also suggest why a non-traditional design
underperforms. The }\textnormal{T}\textnormal{ype C propeller performed
}\textnormal{31}\textnormal{\% worse than the }\textnormal{T}\textnormal{ype }\textnormal{A
3-blade control}\textnormal{. This is likely due to the relatively large tip
vortex apparent on the counterweight in Fig. 3b.}

\textnormal{The testing }\textnormal{setup (}\textnormal{Figure 5) had two main
requirements: the ability to measure frequencies of propeller noise and
}\textnormal{the ability to }\textnormal{measure the thrust produced by the
propellers. To avoid detecting the }\textnormal{propeller wash (noise
created by airflow), }\textnormal{the microphone was stationed on the left
side of the propeller. Furthermore, to ensure that background noise and
reverberation from the environment were not picked up by the microphone,
a sound-dampening chamber was created by lining foam acoustic padding on
the walls of a testing chamber, ensuring that majority of the sound picked
up resulted from the propeller directly {[}9{]}. To measure thrust, a
thrust stand was created by mounting the propeller and motor on a load
cell, which converts forces into }\textnormal{a}\textnormal{ measurable
electrical signal to quantify thrust. Signals were read using an HX711 analog-to-digital converter. The propellers were tested at both
6,000 and 12,000 RPM to ensure an accurate reading in the differences in
frequency and thrust.}

%\textbf{\textnormal{Circuit 1: Arduino PWM ESC: }}\textnormal{To control the
%motor RPM, an Arduino microcontroller was used to output a 5V PWM signal
%to the motor ESC, an off-the-shelf component.}
%
%\textbf{\textnormal{Circuit 2: HX711 Load Cell Circuit:}}\textnormal{ To read
%thrust values, a load cell and HX711 }\textnormal{b}\textnormal{reakout board
%}\textnormal{were used }\textnormal{to make the load cell signal readable via
%Arduino. The data was then }\textnormal{sent to}\textnormal{ a serial data
%terminal on a computer.}

\textnormal{To measure the speed of each propeller, reflective tape was
applied to the blades of the propeller for detection by a laser
tachometer rated up to 100 kRPM.}

\textnormal{All}\textnormal{ propellers were 3D printed using an AnkerMake M5
}\textnormal{printer }\textnormal{with }\textnormal{white }\textnormal{PLA
filament}\textnormal{, using }\textnormal{print settings as detailed in Table
}\textnormal{III}\textnormal{.}

\begin{table}[h]
    \caption{3D Printing Settings}
    \label{tab:3d_printing_settings}
    \centering
    \begin{tblr}{colspec={X[3,l] X[3,l]}, hlines}
        \textbf{Parameter} & \textbf{Value} \\
        Print speed & 250 mm/s \\
        Infill density & 100\% \\
        Layer height & 0.12 mm \\
        Extruder temperature & 205 ℃ \\
        Bed temperature & 55 ℃ \\
        Support type & Grid \\
    \end{tblr}
\end{table}

\textnormal{The microphone used was a Knowles MEMS (microelectronic-mechanical
systems) microphone. The audio signal from
the mic}\textnormal{rophone}\textnormal{ was analyzed using Audacity and decomposed }\textnormal{into individual frequencies using
}\textnormal{the}\textnormal{ Fast Fourier Transform (FFT)}\textnormal{ and}\textnormal{
}\textnormal{converted into }\textnormal{their sound pressure level (SPL)
}\textnormal{in decibels (}\textnormal{dB}\textnormal{) {[}10{]}. SPLs are converted
into power spectral density to visualize how the signal power is
distributed across different frequencies. This method of analysis allows
}\textnormal{the}\textnormal{ pinpoint}\textnormal{ing of}\textnormal{ specific
frequencies where distinct variations}\textnormal{ in intensity}\textnormal{
were noticed when comparing the different propellers in terms of tonal
and broadband noise produced.}

\section{Results}

\begin{table}[h]
    \caption{Thrust (N) by Propeller Design}
    \label{tab:propeller_thrust}
    \centering
    \begin{tblr}{colspec={X[4,l] X[2,c] X[2,c]}, hlines}
        \textbf{Propeller Design} & \textbf{6 kRPM} & \textbf{12 kRPM} \\
        Type A - 3-blade control & \textnormal{0.63} & \textnormal{2.36} \\
        Type B - 3-loop toroidal & \textnormal{0.73} & \textnormal{2.90} \\
        Type C - 2-loop toroidal with uneven blade spacing and counterweight & \textnormal{0.45} & \textnormal{2.09} \\
        Type D - 6-blade control & \textnormal{0.65} & \textnormal{2.71} \\
        Type E - 6-blade with uneven blade spacing & \textnormal{0.63} & \textnormal{2.45} \\
    \end{tblr}
\end{table}

\textnormal{ }\textnormal{Although load cell thrust values (Table
}\textnormal{IV}\textnormal{) differ from the }\textnormal{simulated values (Fig. 
4)}\textnormal{, relative thrust values are similar for types B, D, and E in
both data sets. This shows that there is }\textnormal{ }\textnormal{little to no
disadvantage}\textnormal{, regarding thrust, for the noise-reducing designs
when compared to the conventional designs}\textnormal{. In fact, the
}\textnormal{toroidal propeller (type B) }\textnormal{even outperformed
}\textnormal{the }\textnormal{convention}\textnormal{al designs}\textnormal{.}\textnormal{ }

\textnormal{ }\textnormal{It }\textnormal{was determined that}\textnormal{
propell}\textnormal{e}\textnormal{rs type B, D, and E }\textnormal{consistently
}\textnormal{produced the most thrust}\textnormal{, }\textnormal{so they were
}\textnormal{tested for acoustics.}\textnormal{ Types A and C were not analyzed
for acoustics due to their }\textnormal{lesser}\textnormal{ thrust
performance.}

\textnormal{ The previously outlined acoustic data processing methodology
was used on the audio signals, and the PSD for each propeller was
weighted using the A-weighting function, to better represent the
perceived annoyance by humans. Using the 6-blade conventional propeller,
Type D, as a control, the unconventional propellers were compared. }

\begin{figure}[ht]
    \centering
    \includegraphics[width=\linewidth, keepaspectratio]{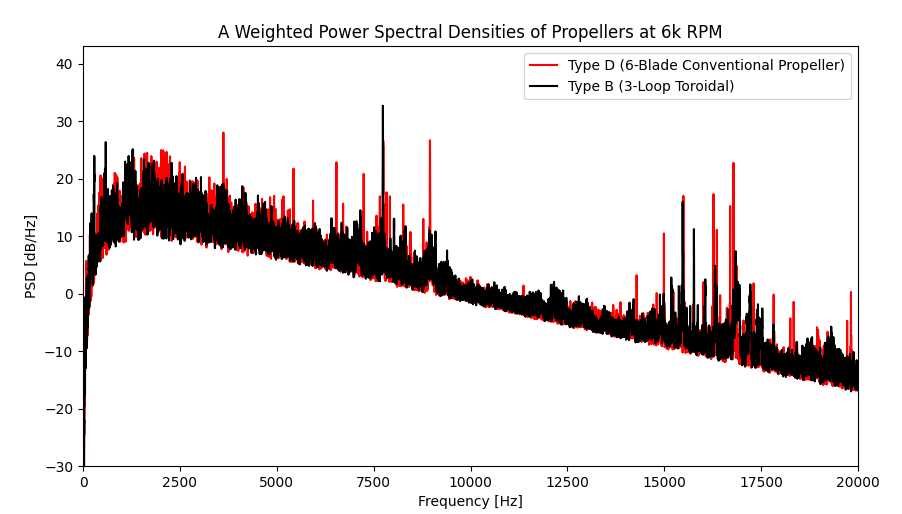}
    \caption{A-weighted power spectral densities for Types D \& B.}
    \label{fig:psd_d_b}
\end{figure}

\begin{figure}[ht]
    \centering
    \includegraphics[width=\linewidth, keepaspectratio]{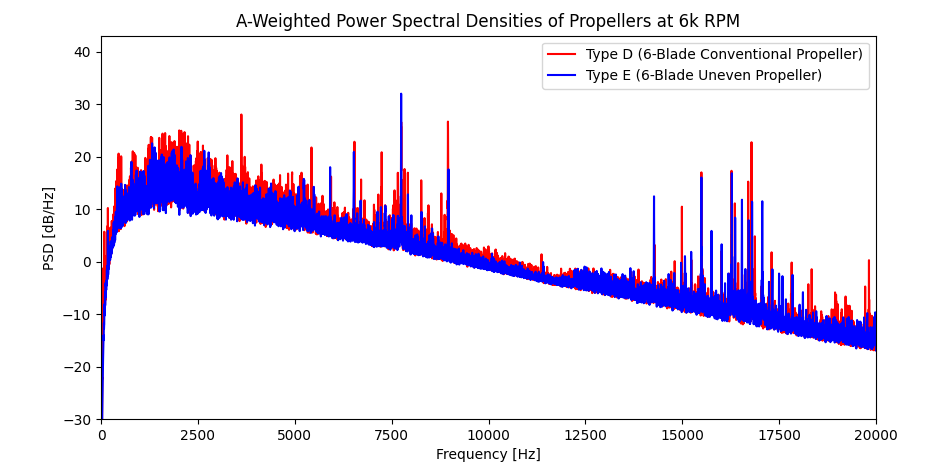}
    \caption{A-weighted power spectral densities for Types D \& E.}
    \label{fig:psd_d_e}
\end{figure}

\textnormal{The decomposition of }\textnormal{the}\textnormal{ recordings showed
several discrete spikes in SPL at multiples of the BPF used to assess
the psychoacoustic properties of the propellers tested.}\textnormal{~}

\textnormal{With Type D as a control}\textnormal{, both
propell}\textnormal{e}\textnormal{rs type B and E had lowered peaks specifically
at higher frequencies. Type B had significantly lower peaks than any
other }\textnormal{propell}\textnormal{e}\textnormal{r}\textnormal{ and}\textnormal{ had the
sound profile with the least }\textnormal{perceived }\textnormal{annoyance
overall. Type E also showed decreased peaks at higher frequencies and
overall, but the difference was not as pronounced as Type B.}

\textnormal{From Fig. 6 and 7 it can be noted that there is an increase in
the number of spikes in PSD at multiples of the BPF in the higher
frequencies of the uneven spaced propeller compared to the evenly spaced
propeller. However, the intensity of these peaks is less pronounced in
the uneven spacing, contributing to previous findings mentioned
earlier}\textnormal{ regarding decreased perceived loudness}\textnormal{. The
overall broadband noise created by the uneven spaced propeller is less
than the noise created by the evenly spaced propeller, showing that
these changes to propeller design have potential benefits in terms of
reducing noise pollution.}

\section{Conclusion}

\textnormal{It is predicted that the analyses of this investigation will
help to accelerate the future of noise pollution-reducing propellers by
promoting the exploration of effective designs that still retain thrust
with innovative geometries such as the toroidal geometry and uneven
spacing of propeller blades. New studies could
}\textnormal{investigate}\textnormal{ creating a calculable method to determine
a propeller spacing with the least amount of high frequency noise
generated}\textnormal{, possibly involving the use of both toroidal geometry
and uneven blade spacing, as both prove to have significant decreases in
perceived loudness within the experimental data.}

\textnormal{As drones become more prevalent in society for their wide range
of use cases, more attention must be given to their effects upon the
environment. The inclusion of techniques to combat noise pollution
caused by these drones is important, as this is not }\textnormal{currently
a}\textnormal{ major consideration. Technologies which are detrimental to
their environments may struggle with wider adoption, therefore
minimizing drone noise pollution is a worthy avenue for future
research}\textnormal{ to ensure the feasibility of drone usage in populated
environments.}

\section{References}

\textnormal{{[}1{]}}\textnormal{ }\textnormal{K. M. Hasan, W. S. Suhaili, S. H. Shah
Newaz and M. S. Ahsan, "Development of an aircraft type portable
autonomous drone for agricultural applications," in }\emph{\textnormal{2020
International Conference on Computer Science and Its Application in
Agriculture (ICOSICA)}}\textnormal{, Bogor, Indonesia, 2020, pp. 1-5.}

\textnormal{{[}2{]}}\textnormal{ }\textnormal{A. Gehlot, R. Singh and D. Singh,
"Modular attachments for several applications with the aid of an aerial
drone," in }\emph{\textnormal{2022 International Interdisciplinary
Humanitarian Conference for Sustainability (IIHC)}}\textnormal{, Bengaluru,
India, 2022, pp. 794-798.}

\textnormal{{[}3{]} "IEEE draft standard for drone applications framework,"
in }\emph{\textnormal{P1936.1, 2020}}\textnormal{ , vol., no., pp.1-29, 26 Feb.
2021.}

\textnormal{{[}4{]} B. Sch}\textnormal{ä}\textnormal{ffer, R. Pieren, K. Heutschi,
J. M. Wunderli, and S. Becker, }\textnormal{``}\textnormal{Drone Noise Emission
Characteristics and Noise Effects on Humans}\textnormal{---}\textnormal{A
Systematic Review,}\textnormal{''}\textnormal{ International Journal of
Environmental Research and Public Health, vol. 18, no. 11, p. 5940, Jun.
2021.}

\textnormal{{[}5{]} D. Steele and S. H. Chon, }\textnormal{``}\textnormal{A
perceptual study of sound annoyance,}\textnormal{''}\textnormal{ in 2nd
Conference on Interaction with Sound, Conference Proceedings, 2007.}

\textnormal{{[}6{]} T. Kim, }\textnormal{``}\textnormal{Reduction of tonal propeller
noise by means of uneven blade spacing,}\textnormal{''}\textnormal{ in UC Irvine
Electronic Theses and Dissertations, UC Irvine, 2016.}

\textnormal{{[}7{]} G. Ion and I. Simion, }\textnormal{``}\textnormal{Performance of
3D printed conventional and toroidal propellers for small multirotor
drones}\textnormal{''}\textnormal{, in Journal of Industrial Design and
Engineering Graphics, vol. 18, no. 1, Jul. 2021.}

\textnormal{{[}8{]} G. Andria, A. Di Nisio, A. M. L. Lanzolla, M.
Spadavecchia, G. Pascazio, F. Antonacci, and G. M. Sorrentino,
}\textnormal{``}\textnormal{Design and performance evaluation of drone
propellers,}\textnormal{''}\textnormal{ in 2018 5th IEEE International Workshop
on Metrology for AeroSpace (MetroAeroSpace), 2018.}

\textnormal{{[}9{]} Caniato M., D\textquotesingle Amore G. K. O., Kaspar,
J., and Gasparella A., }\textnormal{``}\textnormal{Sound absorption performance
of sustainable foam materials: }\textnormal{a}\textnormal{pplication of
analytical and numerical tools for the optimization of forecasting
models,}\textnormal{''}\textnormal{ in Applied Acoustics, vol. 161, Apr. 2020.}

\textnormal{{[}10{]} N. Lenssen, }\textnormal{``}\textnormal{Applications of fourier
analysis to audio signal processing: an investigation of chord detection
algorithms,}\textnormal{''}\textnormal{ in Scholarship @ Claremont, Jan. 2013.}

\end{document}